\documentclass[]{elsart}
\pdfoutput=1

\usepackage{epsfig}
\usepackage{graphics}
\usepackage{graphicx}
\usepackage[centertags]{amsmath}
\usepackage{url}
\usepackage{subfigure}
\usepackage{algorithmic}

\newtheorem{finding}{Finding}

\begin{document}

\begin{frontmatter}

\title{Road planning with slime mould: \\ 
If \emph{Physarum} built motorways it would route M6/M74 through Newcastle}

\author{Andrew Adamatzky and Jeff Jones} 

\address{University of the West of England, Bristol BS16 1QY, United Kingdom \\ 
\{andrew.adamatzky, jeff.jones\}@uwe.ac.uk}

\date{}

\maketitle

\begin{abstract}

\noindent
Plasmodium of \emph{Physarum polycephalum} is a single cell visible by unaided eye. During its foraging behaviour 
the cell spans spatially distributed sources of nutrients with a protoplasmic network. Geometrical structure of the 
protoplasmic networks allows the plasmodium to optimize transfer of nutrients between remote parts of its body, to distributively sense its environment, and make a decentralized decision about further routes of migration. We consider the ten most populated urban areas in United Kingdom and study what would be an optimal layout of transport links between these urban areas from the ``plasmodium's point of view''. We represent geographical locations of urban areas by oat flakes, inoculate
the plasmodium in Greater London area and analyse the plasmodium's foraging behaviour. We simulate the behaviour of the plasmodium using a particle collective which responds to the environmental conditions to construct and minimise transport networks. Results of our scoping experiments show that during its colonization of the experimental space the plasmodium forms a protoplasmic network isomorphic to a network of major motorways except the motorway linking England with Scotland. We also imitate the reaction of transport network to disastrous events and show how the transport network can be reconfigured during natural or artificial cataclysms. The results of the present research lay a basis for future science of bio-inspired urban and road planning.  

\vspace{0.5cm}

\end{abstract}

\end{frontmatter}

\section{Introduction}

Plasmodium of \emph{Physarum polycephalum}\footnote{Order \emph{Physarales}, subclass \emph{Myxogastromycetidae}, class \emph{Myxomecetes}} is a single cell with many diploid nuclei. The plasmodium feeds on microbial creatures and microscopic 
food particles. When colonising its habitat the plasmodium tries to optimize the network of its protoplasmic protrusions and 
tubes to span the maximal number of nutrient sources, and to minimize costs of transportation and intra-cellular communication 
complexity. At the same time the plasmodium positions its body away from the domains of chemo-repellents, low humidity and 
high illumination. Due to its highly sophisticated adaptive behaviour, distributed sensing and actuating,
and decentralized decision making, the plasmodium is considered to be amongst the most prospective experimental prototypes of 
biological amorphous computers~\cite{nakagaki_yamada_1999,nakagaki_2000,nakagaki_2001,nakagaki_2001a,nakagaki_iima_2007}.

Decision-making routines inside the plasmodium's body are executed by interacting 
bio-chemical and excitation waves~\cite{nakagaki_yamada_1999}, redistribution of electrical charges on plasmodium's membrane~\cite{achenbach_1981} and spatio-temporal dynamics of  mechanical waves~\cite{nakagaki_yamada_1999}. The plasmodium's `intelligence' is based on the interactions between propagating patterns. Therefore the plasmodium can be considered as an adaptive 
intelligent reaction-diffusion~\cite{adamatzky_naturewissenschaften_2007} or excitable~\cite{achenbach_1981,adamatzky_bz_trees} 
medium encapsulated in elastic growing membrane --- a novel type of biological computing architecture at the interface of
reaction-diffusion chemical processors~\cite{adamatzky_2005} and amorphous computers~\cite{beal2005}.
Experimental proofs of \emph{P. polycephalum} computational abilities include approximation 
of shortest path~\cite{nakagaki_2001a} and hierarchies of planar proximity graphs~\cite{adamatzky_ppl_2009},
computation of plane tessellations~\cite{shirakawa}, implementation of primitive memory~\cite{saigusa}, 
execution of basic logical computing schemes~\cite{tsuda_2004}, control of robot navigation~\cite{tsuda_2007}, 
and natural implementation of spatial logic and process algebra~\cite{schumann_adamatzky_2009}.

Nature-inspired computing paradigms and experimental implementations have already been successfully applied to the
approximation of a minimal-distance path between two given points of a two-dimensional space or a road network. 
The shortest-path problem is solved in experimental reaction-diffusion chemical systems~\cite{adamatzky_2005},
gas-discharge analog systems~\cite{reyes_2002}, spatially extended crystallization systems~\cite{adamatzky_hotice},
and using computer and mathematical models of collective insects~\cite{dorigo_2004} and 
\emph{P. polycephalum}~\cite{tero_2006}. Intriguing analogies between mycelian fungi networks and road networks are 
discussed in~\cite{jarret_2006}. Previously~\cite{adamatzky_UC07} we have evaluated a road-modeling potential of \emph{P. polycephalum}, however no conclusive results were presented there. In the present paper we made a full-scale --- and as you will 
see further successful --- `attack' on the following \emph{Physarum}-road-building problem. Given major cities represented by 
oat flakes and plasmodium of \emph{P. polycephalum} inoculated in one of the cities, will the plasmodium develop a protoplasmic 
network connecting  oat flake that matches the network of motorways connecting the cities? We show that for UK motorways, connecting ten most populous urban areas, the answer is largely 'yes' with some interesting exclusions. 

The paper is structured as follows. We describe experimental and simulation setup in Sect.~\ref{methods}.  In Sect.~\ref{results} we 
discuss experimental and simulation results on developing transport links between major urban areas by plasmodium. Scoping experiments
on plasmodium imitating reconfiguration of road links and mass migration caused by disastrous accidents in selected 
urban areas are reviewed in Sect.~\ref{disasters}. Section~\ref{discussion} suggest lines of further experiments and 
highlights possible applications of plasmodium-computing.

\section{Methods}
\label{methods}

Plasmodium of \emph{ P. polycephalum} is cultivated in plastic container, on paper kitchen towels sprinkled with 
distilled water and fed with oat flakes\footnote{Asda's Smart Price Porridge Oats}. For experiments we use 
$120 \times 120$~mm polyestyrene square and $90 mm$ diameter round Petri dishes. We use either 2\% agar gel (Select agar, Sigma Aldrich) or a moistured filter paper as a non-nutrient growth substrate. Agar plates and filter papers are cut in a shape of United Kingdom island.

\begin{figure}
\centering
\subfigure[]{\includegraphics[width=0.49\textwidth]{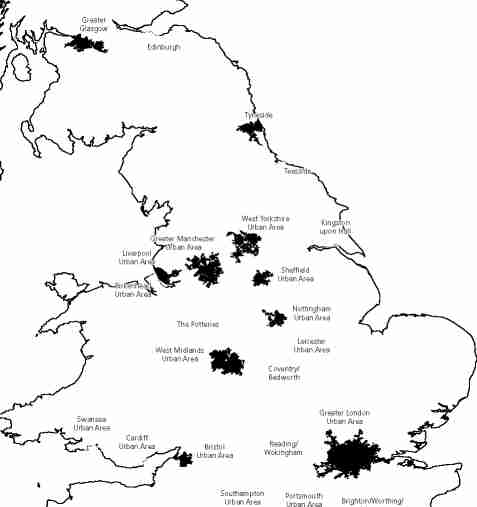}}
\subfigure[]{\includegraphics[width=0.49\textwidth]{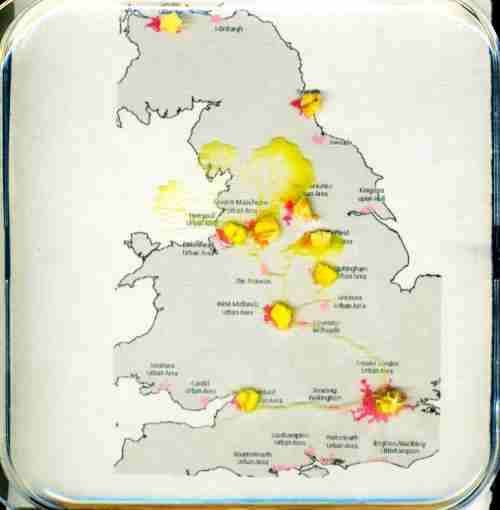}}
\caption{Experimental basics. 
(a)~Schematic map of ten most populous urban areas, shown in black.
(b)~Typical experimental setup: urban areas are represented by oat flakes, plasmodium 
is inoculated in London, the plasmodium spans oat flakes by protoplasmic transport 
network. 
Map of urban areas is adapted from~\cite{pointer_2005}. }
\label{urbanareas}
\end{figure}

We consider the ten most populous urban areas in United Kingdom (Fig.~\ref{urbanareas}a) --- 
Greater London, Bristol, Sheffield, Nottingham, 
Liverpool, Tyneside, Greater Glasgow, West Yorkshire, Greater Manchester,
and West Midlands --- as per 2001 Census\footnote{Office for National Statistics, General Register Office 
for Scotland and Northern Ireland Statistics and Research Agency}, see details and boundaries of the areas 
in~\cite{pointer_2005}. The areas are projected onto gel or filter paper and oat flakes (of size approximately 
matching size and shape of the areas) are placed in the positions of the urban areas (Fig.~\ref{urbanareas}b).
At the beginning of each experiment plasmodium is inoculated in the centre of Greater London Urban Area. 

The Petri dishes with substrate and plasmodium are kept in darkness, at temperature 22-25~C$^o$, except for 
observation and image recording. Periodically the dishes are scanned using Epson Perfection 4490 scanner.
Large-scale disasters in urban areas were imitated by salt 
crystals\footnote{Saxa corse grain sea salt, RHM Foods, CW10 0HD, UK} 
being placed in centres of the urban areas. 

Scanned images of dishes are enhanced for higher visibility, saturation increased to 204, and contrast to 40.
To ease readability of experimental images we provide complementary binary version of each image. The binarization
is done as follows. Each pixel of a color image is assigned black color if red $R$ and green $G$ components of its
RGB color exceed some specified thresholds, $R > \theta_R$, $G > \theta_G$ and blue component $B$ does 
not exceed some threshold value $B < \theta_B$; otherwise, 
the pixel is assigned white color (exact values of the thresholds are indicated in the figure captions as 
$\Theta=(\theta_R, \theta_G, \theta_B)$.  

The computational approximation of \emph{Physarum} is provided by the particle collective approach introduced in ~\cite{jones_transportnet_2008} where a population of very simple mobile particles with chemotaxis-like sensory behaviour were used to construct and minimise spatially represented emergent transport networks in a diffusive environment. The particle approximation corresponds to a particle approximation of LALI (Local Activation Long-range Inhibition) reaction-diffusion pattern formation processes ~\cite{jones_passiveactive_2009} and exhibits a complex range of patterning by varying particle sensory parameters ~\cite{jones_alife_2008}. We assume that each particle in the collective represents a hypothetical unit of \emph{Physarum} plasmodium gel/sol interaction which includes the effect of chemoattractant gradients on the plasmodium membrane (sensory behaviour) and the flow of protoplasmic sol within the plasmodium (motor behaviour). The summation of particle positions corresponds to a static snapshot of network structure whilst the collective movement of the particles in the network corresponds to protoplasmic flow within the network.

Although the model is very simple in its assumptions and implementation it is capable of reproducing some of the spontaneous network formation, network foraging, oscillatory behaviour, bi-directional shuttle streaming, and network adaptation seen in \emph{Physarum} using only simple, local functionality to generate the emergent behaviours. Details of the particle morphology, sensory and motor behavioural algorithms can be found in ~\cite{jones_uc09} and in this paper we use an extension of the basic model (without utilising oscillatory behaviour) to include plasmodium growth and adaptation (growth and shrinkage of the collective). 

Growth and adaptation of the particle model population is currently implemented using a simple method based upon local measures of space availability (growth) and overcrowding (adaptation, or shrinkage, by population reduction). This is undoubtedly a gross simplification of the complex factors involved in growth and adaptation of the real organism (such as metabolic influences, nutrient concentration, waste concentration, slime capsule coverage, bacterial contamination etc.). However the simplification renders the population growth and adaptation more computationally tractable and the specific parameters governing growth and shrinkage are at least loosely based upon real environmental constraints. Growth and shrinkage states are iterated separately for each particle and the results for each particle are indicated by tagging Boolean values to the particles. The growth and shrinkage tests were executed every two scheduler steps and the method employed is specified as follows.

If there are 1 to 10 particles in a $9 \times 9$ neighbouhood of a particle, and the particle has moved forwards successfully, the particle  attempts to divide into two if there is an empty location in the immediate neighbourhood surrounding the particle. If there are 
0 to 24 particles in a $5 \times 5$ neigbourhood of a particle the particle survives, otherwise it is annihilated. 



To replicate the experimental spatial configuration the environment is represented by a colour coded image. Urban areas are indicated by black areas, uninhabitable coastal boundary of the UK mainland is represented by the light grey border region and the background (empty substrate or nutrient-rich substrate, depending on the experiment) is shown as white. We simulate three environmental conditions. The first is a nutrient rich-substrate with background nutrients in addition to oat flakes, corresponding to \emph{Physarum} growth on oatmeal agar. 
The second condition is a high-concentration nutrient source on a non-nutritious substrate. This corresponds to growth on plain agar. 
The third condition is foraging behaviours which approximates dendritic foraging growth on damp filter paper. 

The three different environmental conditions are approximated by adjusting simulated chemoattractant projection strengths and diffusion distances in the particle model. Oat flake positions were represented by projection of chemoattractant to the diffusion map at the locations of urban areas. The projection weight (altering the strength of chemoattractant projection, higher weight values result in more chemoattractant projection to the environment) applied to the urban areas was 0.01 in the nutrient-rich and foraging experiments, and was set to 0.5 for the high-concentration experiments. Diffusion damping (affecting the distance of chemoattractant gradient diffusion, lower values result in longer diffusion distances) was set to 0.1 for nutrient-rich substrate and foraging experiments, and to 0.01 for the high concentration experiments. 

The diffusion 
kernel size was $7 \times 7$ pixels for all experiments. We assume that diffusion of chemoattractant from any urban area is suppressed when the area is covered by particles. The suppression is implemented by checking each pixel within an urban area and reducing the projection value, or concentration of chemoattractants, by multiplying it by 0.001
if there is a particle within a $3 \times 3$ neighbourhood surrounding the pixel. 

Contamination of an urban area with salt was implemented by deleting any particles which entered the contaminated area. Inoculation of a small initial particle population (5 particles) was initiated at random locations within the London urban area. 

Particle sensor offset was 5 pixels and angle of rotation set to 45 degrees. Particle sensor angle was 30 degrees in the foraging experiments (the relatively narrow sensor angle stimulates branching transport networks) and 45 degrees in all other experiments.

 The emergent transport networks are represented in the results by the configuration of the particle collective, shown as a spatial map of particle positions. Where appropriate, superpositions of previous particle occupancy history are shown in grey, and darker shading indicates more frequent transport. Video recordings of the dynamical growth and adaptation of the particle model can be found at \url{http://uncomp.uwe.ac.uk/jeff/urban/urban.htm}.

\section{Development of transport links}
\label{results}

\begin{figure}
\centering
\subfigure[$t=$12~h]{\includegraphics[width=0.32\textwidth]{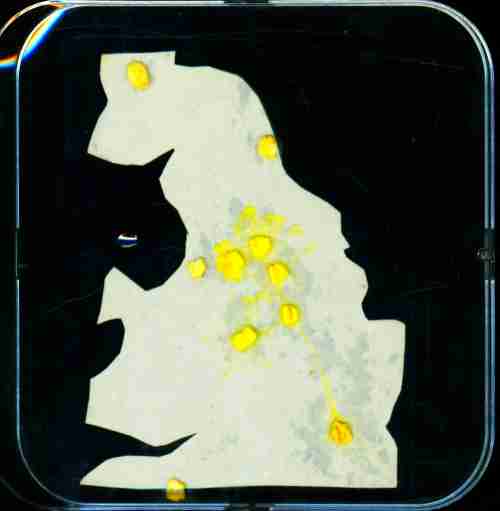}}
\subfigure[$t=$23~h]{\includegraphics[width=0.32\textwidth]{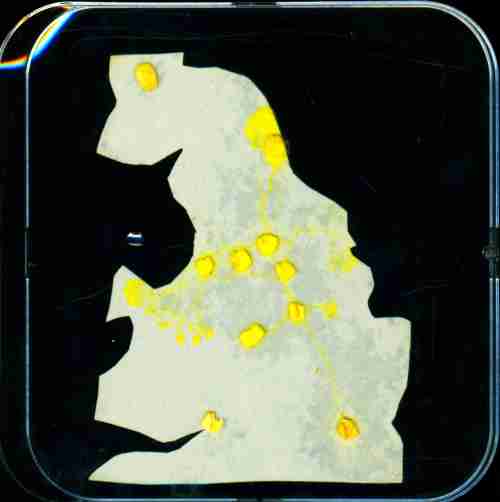}}
\subfigure[$t=$34~h]{\includegraphics[width=0.32\textwidth]{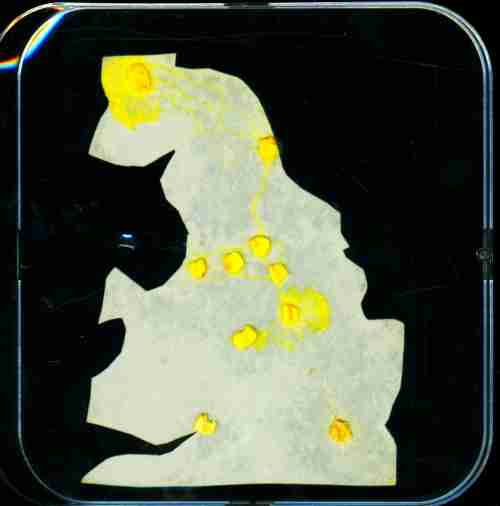}}
\subfigure[$t=$47~h]{\includegraphics[width=0.32\textwidth]{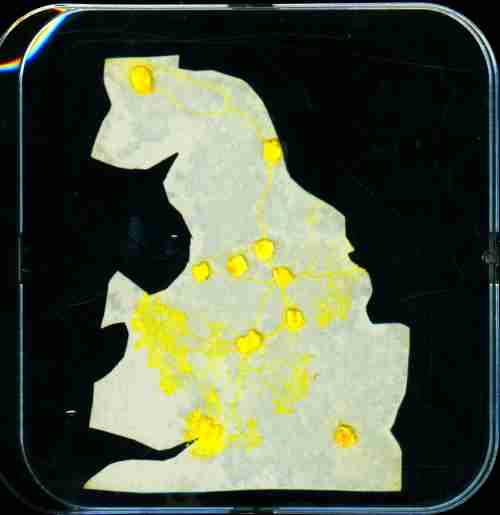}}
\subfigure[$t=$69~h]{\includegraphics[width=0.32\textwidth]{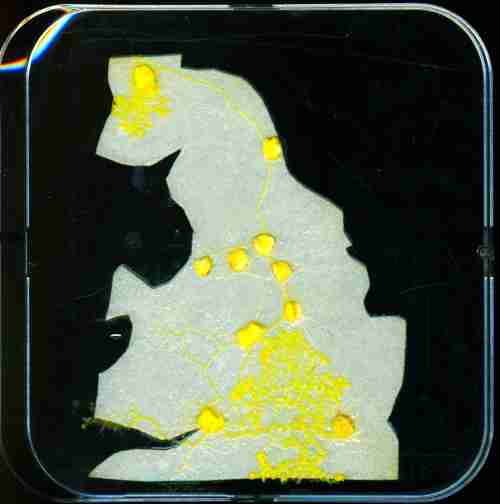}}
\subfigure[$t=$80~h]{\includegraphics[width=0.32\textwidth]{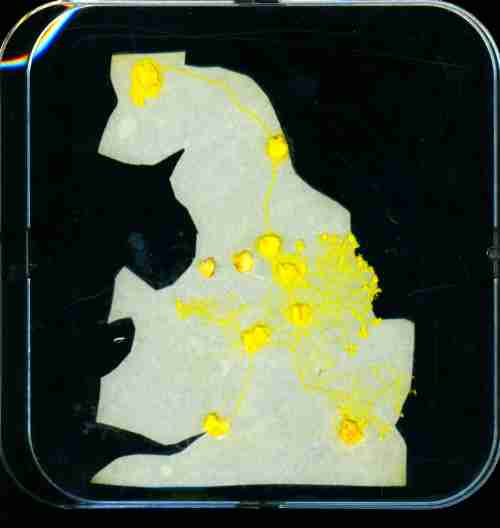}}
\subfigure[$t=$12~h]{\includegraphics[width=0.32\textwidth]{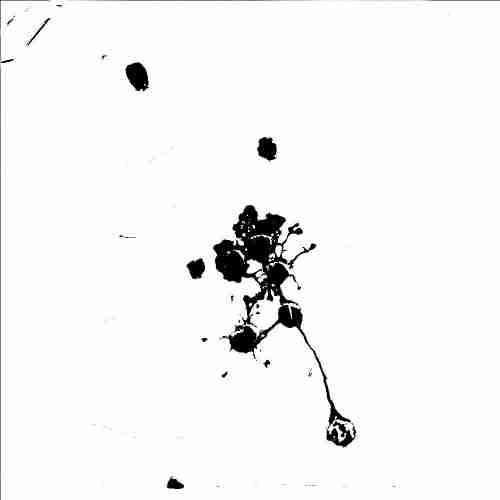}}
\subfigure[$t=$23~h]{\includegraphics[width=0.32\textwidth]{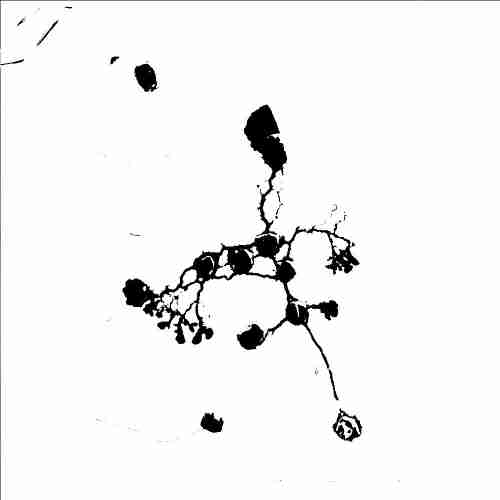}}
\subfigure[$t=$34~h]{\includegraphics[width=0.32\textwidth]{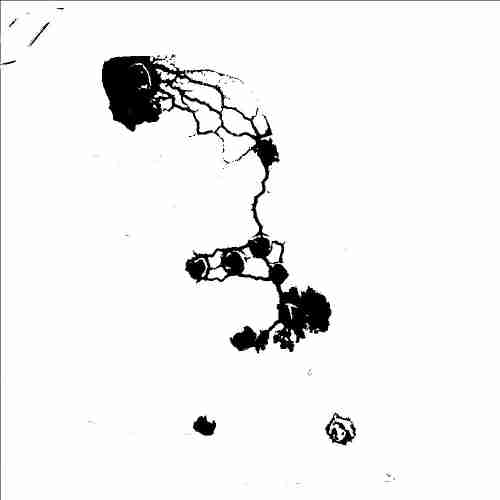}}
\subfigure[$t=$47~h]{\includegraphics[width=0.32\textwidth]{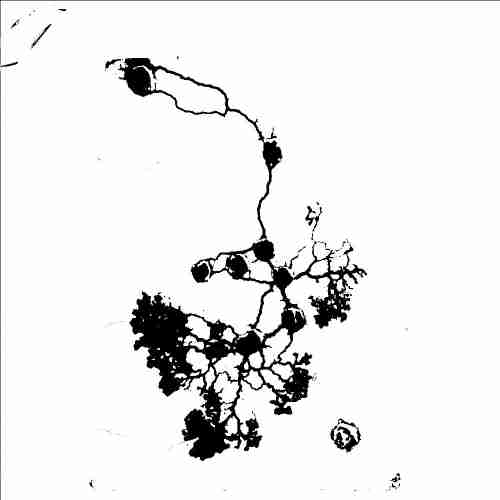}}
\subfigure[$t=$69~h]{\includegraphics[width=0.32\textwidth]{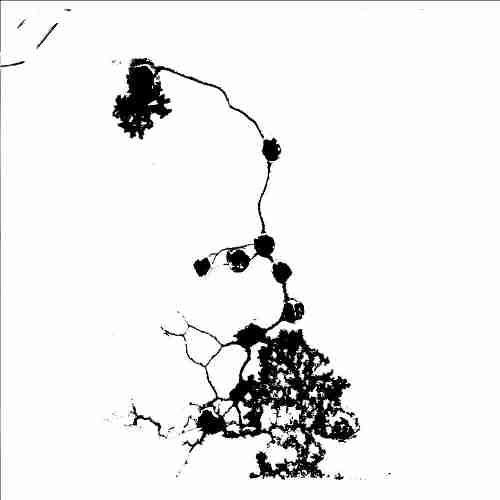}}
\subfigure[$t=$80~h]{\includegraphics[width=0.32\textwidth]{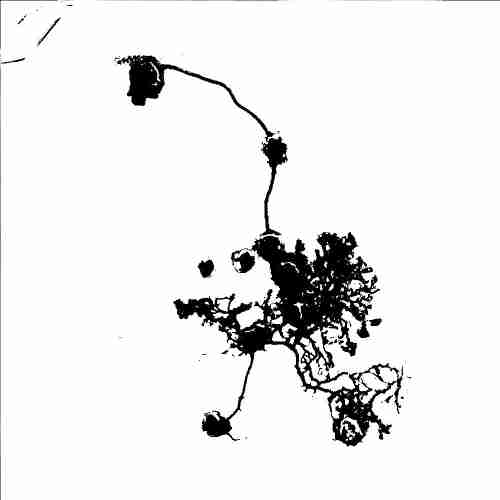}}
\caption{Typical plasmodium development:
(a)--(f)~scanned image of experimental Petri dish;
(g)--(l)~binarized images, $\Theta=(200,200,150)$}
\label{frh1}
\end{figure}

\begin{figure}
\centering
\subfigure[$t=$63~h]{\includegraphics[width=0.32\textwidth]{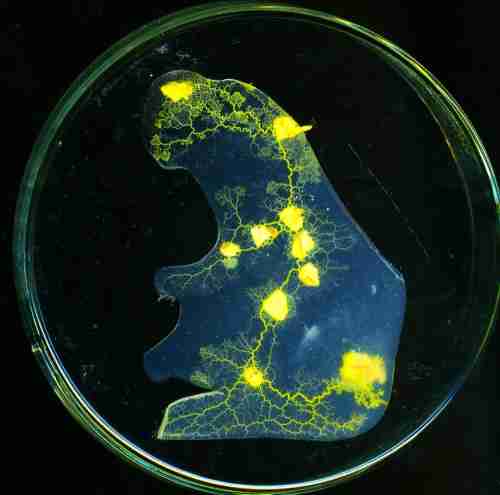}}
\subfigure[$t=$63~h]{\includegraphics[width=0.32\textwidth]{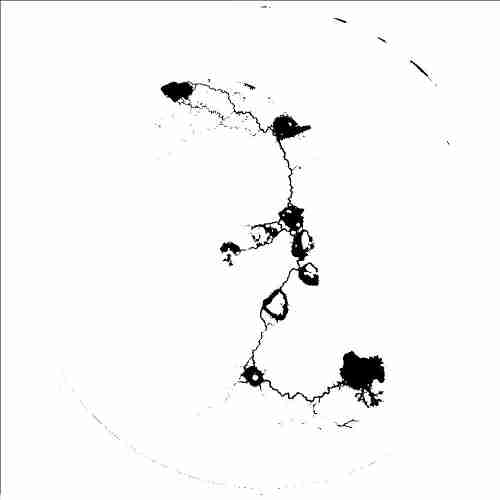}}
\subfigure[$t=$47~h]{\includegraphics[width=0.32\textwidth]{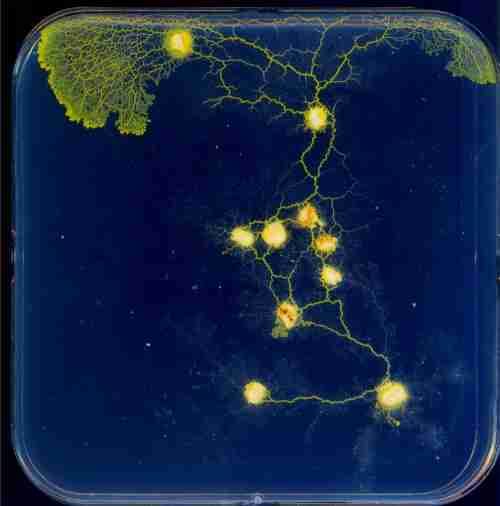}}
\subfigure[$t=$47~h]{\includegraphics[width=0.32\textwidth]{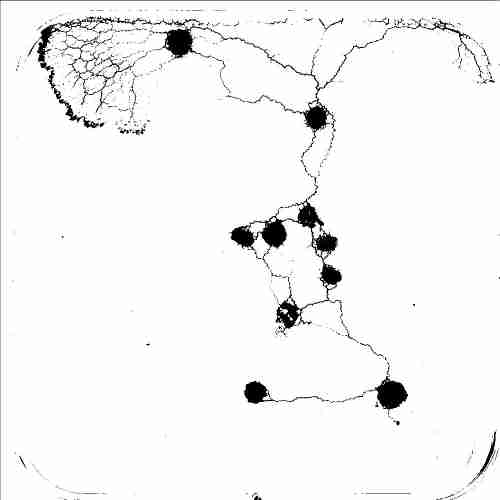}}
\subfigure[$t=$53~h]{\includegraphics[width=0.32\textwidth]{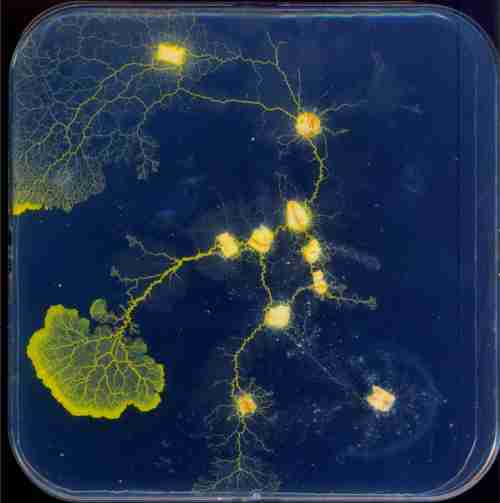}}
\subfigure[$t=$53~h]{\includegraphics[width=0.32\textwidth]{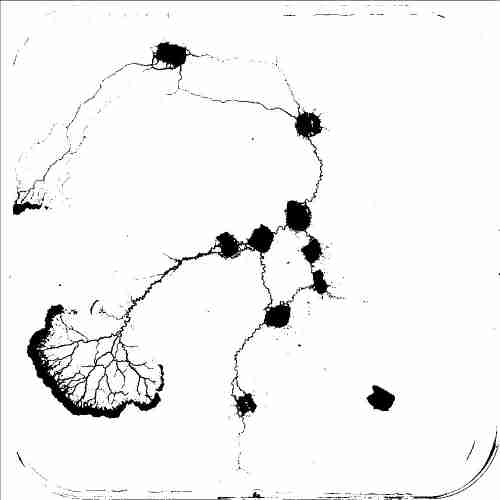}}
\subfigure[$t=$46~h]{\includegraphics[width=0.32\textwidth]{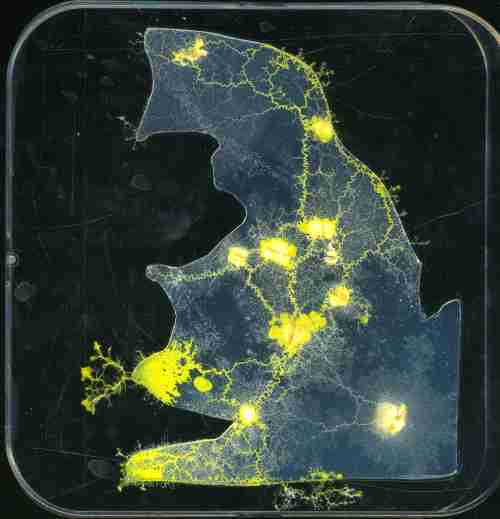}}
\subfigure[$t=$46~h]{\includegraphics[width=0.32\textwidth]{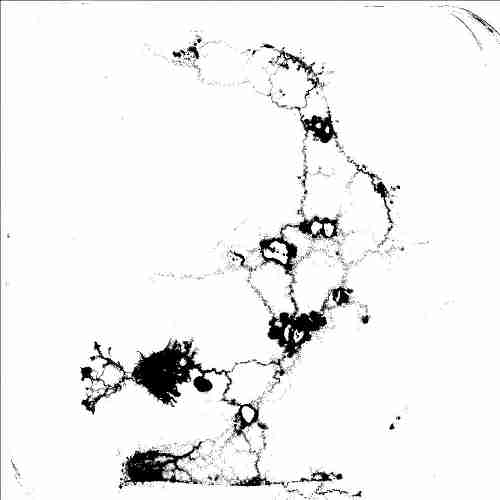}}
\caption{Example of protoplasmic networks which occurred in experiments. Results of four independent
experiments are shown in (ab), (cd), (ef) and (gh), respectively. Time passed after inoculation of 
plasmodium in London is shown in sub-figure caption. (a), (c), (e), (g) are scanned image 
of experimental Petri dishes; (b), (d), (f), (h) are binarized images. Thresholds of binarization
are $\Theta=(170,170,100)$ for (b) and (d), and $\Theta=(130,130,100)$ for (f) and (h).}
\label{examples}
\end{figure}

Typically, being placed in the centre of London urban area, plasmodium consumes some 
nutrients from its nearest (London) oat flake and starts propagating North, North-West or West
(Fig.~\ref{frh1}ag). Birmingham (Fig.~\ref{frh1}ag) and Bristol (Fig.~\ref{examples}cd)
are the usual candidates which are spanned by London-originated plasmodium. When urban area areas concentrated in Midlands
are colonised by plasmodium and linked by protoplasmic tubes the plasmodium heads North towards 
the Tyneside urban area (Fig.~\ref{frh1}bh). After taking on Tyneside the plasmodium propagates North, cross Scottish boundaries
and finally reaches Glasgow urban area (Fig.~\ref{frh1}ci). Then plasmodium continues colonization of the substrate until
all urban areas (sources of nutrients) are colonized (Fig.~\ref{frh1}d--f and Fig.~\ref{frh1}j--l). 

Example of plasmodium networks connecting urban areas are shown in Fig.~\ref{examples}. The figures demonstrate
that, in general, the structure of the network does not depend significantly on size and shape of the substrate but mainly on 
the configuration of sources of nutrients: 90~mm round Petri dish (Fig.~\ref{examples}ab), 120~mm rectangular dishes fully covered with 
agar gel (Fig.~\ref{examples}c--f) and shape of UK island cut of agar gel plate in 120~mm rectangular 
Petri dish (Fig.~\ref{examples}gh). Noticeably, the plasmodium does not stop its foraging activity even when all sources
of nutrients are colonised. It propagates away from the `designated' area (Fig.~\ref{examples}c--f) unless stopped by 
an unfriendly substrate (like the bottom of a plastic Petri dish not covered by gel, Fig.~\ref{examples}gh).

\begin{figure}
\centering
\subfigure[$t=$8~h]{\includegraphics[width=0.32\textwidth]{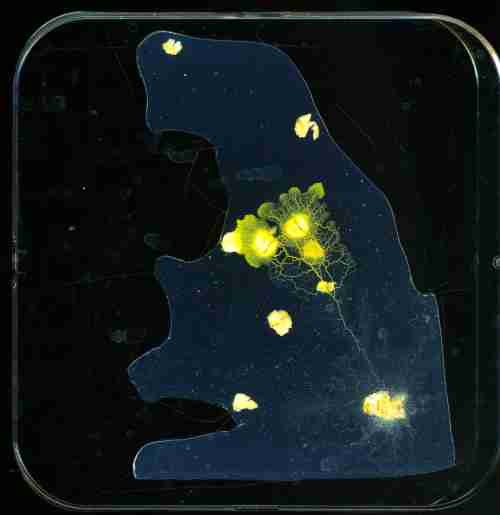}}
\subfigure[$t=$31~h]{\includegraphics[width=0.32\textwidth]{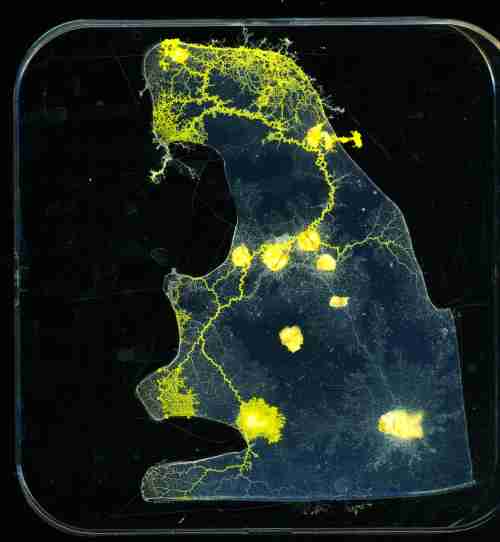}}\\
\subfigure[$t=$8~h]{\includegraphics[width=0.32\textwidth]{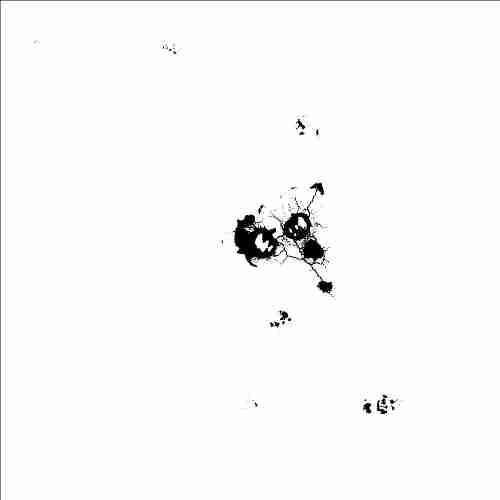}}
\subfigure[$t=$31~h]{\includegraphics[width=0.32\textwidth]{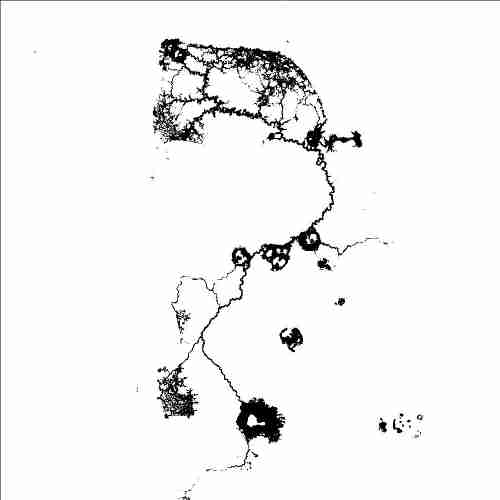}}
\caption{Examples of reconfigurations of protoplasmic network:
(a) and (b)~scanned images, (c) and (d)~binarised images, $\Theta=(140,160,100)$}
\label{r5}
\end{figure}

The plasmodium does not always keep all sources of nutrients spanned by its protoplasmic tubes. Sometimes some tubes
cab be abandoned during colonization. Thus in Fig.~\ref{r5} we see that at the beginning of its development
plasmodium links London and Nottingham (Fig.~\ref{r5}ac). When urban areas in Midlands are colonised and linked
by protoplasmic tubes to Tyneside and Glasgow areas, the plasmodium abandons its tube connecting Greater London
and  Nottigham urban areas (Fig.~\ref{r5}bd).

\begin{figure}
\centering
\subfigure[$t=$ 111, 186, 289, 377, 836, 1864, 4020 and 9992 scheduler steps]{\includegraphics[width=0.9\textwidth]{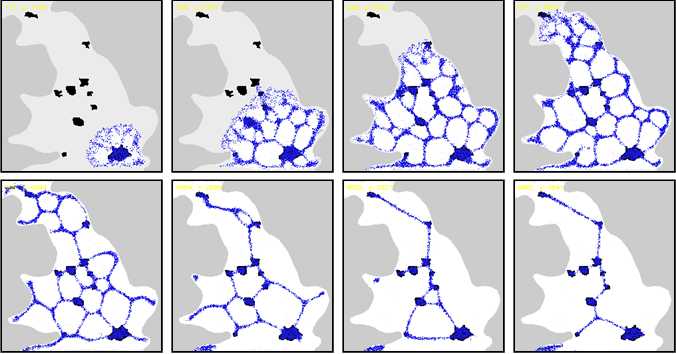}}
\subfigure[$t=$  420, 527, 1534 and 2999 scheduler steps]{\includegraphics[width=0.9\textwidth]{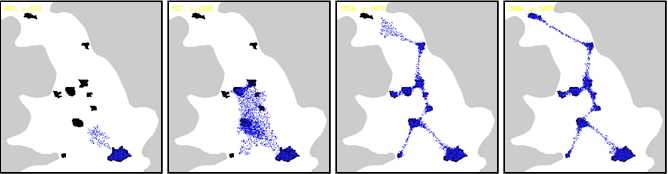}}
\subfigure[$t=$ 2340, 7736, 11780, 13068, 22552 and 29412 scheduler steps]{\includegraphics[width=0.7\textwidth]{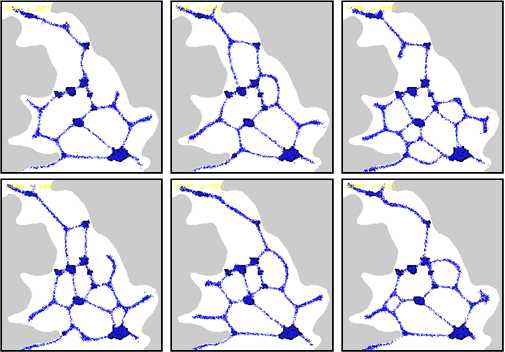}}
\caption{Particle approximations of emergent transport networks with different nutrient concentrations, diffusion properties and foraging behaviour.
(a) Evolution of radial expansive growth front and subsequent network adaptation on a nutrient-rich substrate, (b) evolution of network with a propagating growth front in a highly diffusive environment, (c) snapshots showing dynamical reconfiguring network when foraging behaviour is used. }
\label{model_growth}
\end{figure}

The particle approximation results varied depending on the nutrient and diffusion conditions of the simulated environment. Fig.~\ref{model_growth} (a) illustrates radial expansive growth in a nutrient-rich substrate environment, approximating the growth on oatmeal agar, for example. The background nutrients (very pale grey on areas not reached by population) are assumed to be consumed immediately on contact whereas the the simulated oat flakes (urban areas) remain a persistent source of nutrients. The expansion of the population continues until all the substrate area has been convered and then the particle population automatically adapts and shrinks the transport network to encompass and connect the urban areas. Fig.~\ref{model_growth} (b) is in an environment with a nutrient poor substrate but high concentration of chemoattractants emitted from the urban areas in highly diffusive conditions. The population expands to locate the nearest source of food and then continues until all food sources are located and the network between them is minimised. Although this behaviour appears to be computationally efficient, it does not reflect the actual reconfiguration behaviour seen in \emph{Physarum} and this is more accurately reflected in Fig.~\ref{model_growth} (c) where a narrow sensor angle of 30 is used to imitate foraging, branching behaviour. The discovery of the food-source nodes (not shown) is not as predictable or direct when nutrient conditions are poor and the transport network undergoes constant reconfiguration whilst keeping the food sources covered.

Even a limited number of examples (e.g. Figs.~\ref{examples} and \ref{r5}) demonstrate that plasmodium of 
\emph{P. polycephalum} is a very dynamic system, in which the morphology is continuously changing and in which 
spatio-temporal dynamics rarely reaches a fixed stable point (unless humidity decreases and plasmodium 
forms a sclerotium). There is no such thing as a stationary configuration of protoplasmic network, therefore when 
extracting a generalised graph of transport links from plasmodium experiments we can only build a 
\emph{Physarum}-graph $\mathbf P$ as follows. 

Let $\mathbf U$ be a set of ten most populous urban areas, $\mathbf S$=$\{ S_1, \cdots, S_k \}$ is a set
of series $S_i$, $i=1, \ldots, k$ ($k$ is a number of experiments), of scanned images of plasmodium 
networks,  $S_i = (s^1_i, \ldots, s^{m_i}_i)$.  For any two areas $a$ and $b$ from $\mathbf U$ 
the weight of edge $(ab)$ is calculated as follows: $w(ab) = \sum_{S_i \in \mathbf S} \chi(S_i, a, b)$, where $\chi(S_i, a, b) = 1$ if there is at least one snapshot $s \in S_i$ which shows a protoplasmic tube connecting $a$ and $b$.  
We do not take into account exact  configuration of the protoplasmic tubes but merely their existence.  
Each protoplasmic tube is counted just once for any particular experiment. We will also consider sub-graphs
 $\mathbf P_{\alpha}$ of $\mathbf P$, $\alpha= 5, 10, 12$ defined as follows: 
 for $a, b \in \mathbf U$: $(ab) \in \mathbf P_\alpha$ if $w(ab) > \alpha$.

\begin{figure}
\centering
\subfigure[]{\includegraphics[width=0.49\textwidth]{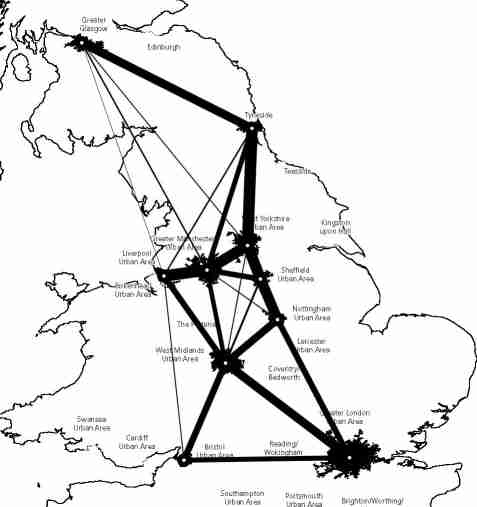}}
\subfigure[]{\includegraphics[width=0.49\textwidth]{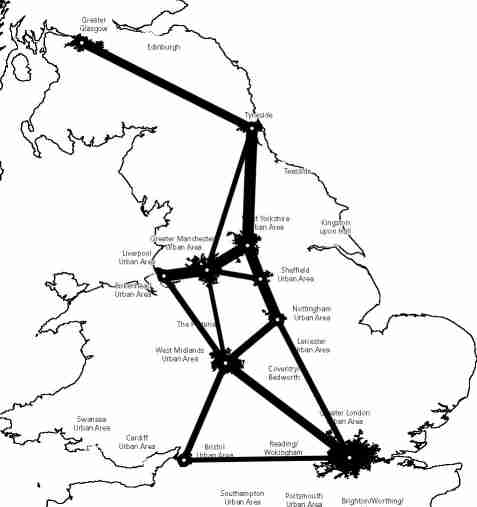}}
\subfigure[]{\includegraphics[width=0.49\textwidth]{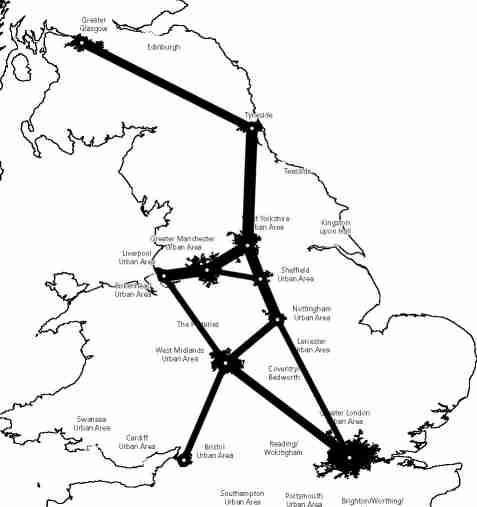}}
\subfigure[]{\includegraphics[width=0.49\textwidth]{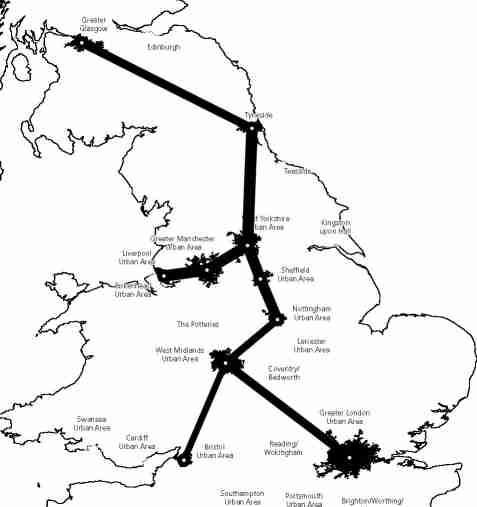}}
\caption{\emph{Physarum}-graphs for various values of edge weights: 
(a)~all edges of Physarum graph are shown, thickness of each edge is proportional to the edge's weight, 
(a)--(d) only edges with weights exceeding 5~(b), 10~(c) and 12~(d) are shown. }
\label{Fgraphs}
\end{figure}

\emph{Physarum}-graphs extracted from 25 laboratory experiments are shown in Fig.~\ref{Fgraphs}, maximum edge weight is 22. 
The graph becomes planar when we remove edges with weights below 6 (Fig.~\ref{Fgraphs}b). The graphs is acyclic, or a tree,
when only edges appearing in over 40\% of experiments are shown (Fig.~\ref{Fgraphs}d). If we increase cut-off value to 14 the graph becomes disconnected, and the node corresponding to the Bristol urban area becomes isolated.

\begin{figure}
\centering
\subfigure[]{\includegraphics[width=0.49\textwidth]{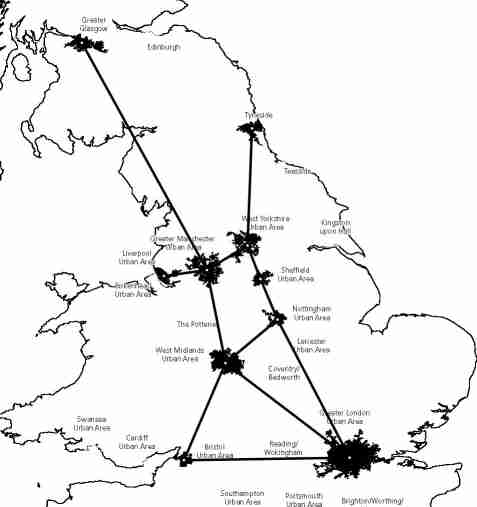}}
\caption{Graph $\mathbf M$ of man-made motorway network connecting ten most populous urban areas.}
\label{motorways}
\end{figure}

Let us check if there  is any correspondence between transport links built by \emph{Physarum} and man-build motorways. 
We construct the motorway graph $\mathbf M$ as follows\footnote{Graph $\mathbf M$ is extracted from 
motorway network as shown in \url{maps.google.com} and \url{www.openstreetmap.org}}. 
Let $\mathbf U$ be a set of ten most populous urban areas, for any two areas $a$ and $b$ from $\mathbf U$, 
the nodes $a$ and $b$ are connected by an edge $(ab)$ if there is a motorway starting in 
vicinity of $a$ and passing in vicinity of $b$ and not passing in vicinity of any other 
urban area $c \in \mathbf U$. The motorway graph $\mathbf M$ is shown in Fig.~\ref{motorways}. 
By comparing $\mathbf M$ (Fig.~\ref{motorways}) and \emph{Physarum} graphs $\mathbf P$ and its 
subgraphs (Fig.~\ref{Fgraphs}) we found that.

\begin{finding}
Motorway graph $\mathbf M$ is a sub-graph of \emph{Physarum}-graph $\mathbf P$. 
\end{finding}

This shows that --- in princple --- the distributed `logic' underpinning plasmodium's decision-making 
routines corresponds to human logic behind road-planning decisions. However graph $\mathbf P$ 
is not planar thus poses no practical importance. Motorway link M6/M74 connecting Great Manchester and Great Glasgow urban area is represented 
by plasmodium only in 3 of 25 experiments. The corresponding edge does not appear in graphs $\mathbf{P}_5$, 
$\mathbf{P}_{10}$ and $\mathbf{P}_{12}$ (Fig.~\ref{Fgraphs}b--d).   Motorway M4, linking Greater London and Bristol urban
areas, is represented by \emph{Physarum} only in 20\% of experiments, graph $\mathbf{P}_5$ in Fig.~\ref{Fgraphs}b. 

\begin{finding}
\emph{Physarum polycephalum} satisfactory approximates motorway network linking the ten most populous urban areas in United Kingdom except 
the motorway link M6/M74 connecting Greater Manchester and Greater Glasgow urban areas. 
\end{finding}

\begin{figure}
\centering
\subfigure[]{\includegraphics[width=0.48\textwidth]{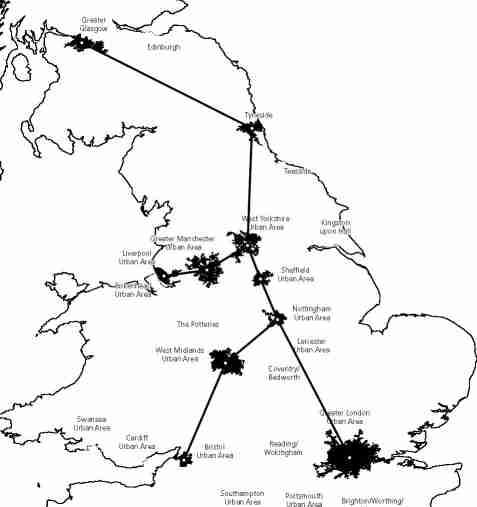}}
\subfigure[]{\includegraphics[width=0.48\textwidth]{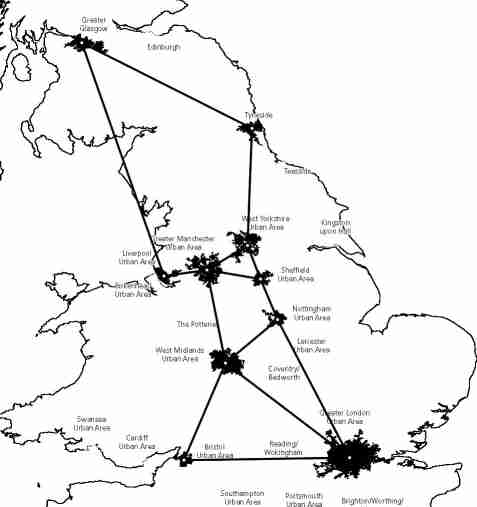}}
\caption{Two proximity constructed on urban areas $\mathbf U$: 
(a)~relative neighbourhood graph $\mathbf{RNG}$,
(b)~Gabriel graph $\mathbf{GG}$.}
\label{proximity}
\end{figure}

Is there a rationale behind plasmodium's behaviour? Let us have a look at two most popular planar proximity graphs, Relative Neighbourhood graph~\cite{toussaint_1980} $\mathbf{RNG}$ (Fig.~\ref{proximity}a) and Gabriel graph~\cite{gabriel_sokal_1969, matula_sokal_1984} $\mathbf{GG}$ (Fig.~\ref{proximity}b) constructed over nodes corresponding to centres of urban areas. 
Points $a$ and $b$ are connected by an edge in $\mathbf{RNG}$ if no other point $c$ is closer to $a$ and $b$ than 
$dist(a,b)$~\cite{toussaint_1980}. Points $a$ and $b$ are connected by edge in $\mathbf{GG}$ is disc with diameter $dist(a,b)$ 
centered in middle of the segment $ab$ is empty~\cite{gabriel_sokal_1969, matula_sokal_1984}.
The graphs are related as  
$\mathbf{RNG} \subseteq \mathbf{GG}$~\cite{toussaint_1980,matula_sokal_1984,jaromczyk_toussaint_1992}. Both graphs are imporant
in spatial analysis and statistics, particularly $\mathbf{GG}$ which was invented specially to analyse and simulate 
geographical distribution of biological populations~\cite{gabriel_sokal_1969}. Our experiments
shown that:

\begin{finding}
$\mathbf{P}_{10} \subset \mathbf{GG}$ and $\mathbf{P}_{12} = \mathbf{RNG}$.
\end{finding}

Moreover $\mathbf{P}_{12}$ is a minimum spanning tree $\mathbf{MST}$ over $\mathbf U$. 
We know that $\mathbf{MST} \subseteq \mathbf{RNG}$~\cite{toussaint_1980} but in the particular case of 
urban areas' configurations $\mathbf U$ we even have $\mathbf{MST}({\mathbf U}) = \mathbf{RNG}({\mathbf U})$.

With regards to the relationship between motorway graph and proximity graphs we see that 
$\mathbf M $ is neither sub- nor super-graph of $\mathbf{RNG}$ and $\mathbf{GG}$. To transform $\mathbf M$ to $\mathbf{RNG}$
one  needs to remove two edges from and relocate one edge in $\mathbf M$, while to transform $\mathbf M$ to $\mathbf{GG}$ it 
no edges should be removed from but three edges added to $\mathbf M$. The edge connecting Tyneside and Greater Glasgow is present in 
$\mathbf{RNG}$ and $\mathbf{GG}$ but absent in $\mathbf M$.

\begin{finding}
Experiments with plasmodium of \emph{Physarum polycephalum} show that motorway M6/M74 is not optimally positioned and 
should be rerouted from Newcastle to Glasgow. Alternatively M6/M74 may remain intact but new motorway Newcastle-Glasgow must be built.
\end{finding}

The evolution and minimisation of transport networks by the particle approximation in diffusive, nutrient-rich environments closely approximates certain variants of the \emph{Physarum}-graphs, most notably $\mathbf{P}_{12}$. Another similarity between \emph{Physarum}-graphs and the emergent transport networks is the common absence of the Bristol-London link and the Manchester-Glasgow link. The connectivity of the transport networks is also affected by the nutrient and diffusion strength --- both the nutrient-rich substrate example and high-concentration example show an approximation of a tree structure similar to the $\mathbf{RNG}$. With the highly foraging example (Fig.~\ref{model_growth}c) the connectivity is more varied and dynamic but there is still the tendency to avoid the Manchester-Glasgow and Bristol-London links.

\begin{figure}
\centering
\subfigure[]{\includegraphics[width=0.3\textwidth]{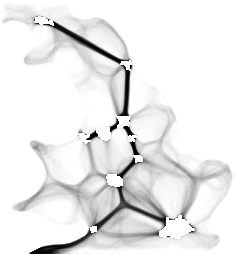}}
\subfigure[]{\includegraphics[width=0.3\textwidth]{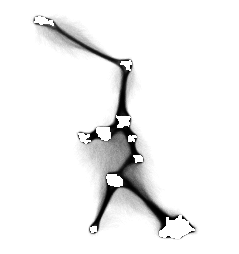}}
\subfigure[]{\includegraphics[width=0.3\textwidth]{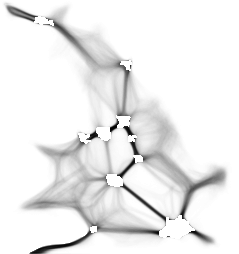}}
\caption{Superpositions of the history of emergent transport networks produced by the particle populations. Darker shading indicates more frequently visited --- more stable --- paths. (a) Superposition of network evolution on a nutrient-rich substrate, (b) Superposition of evolution in a highly diffusive environment, (c) Superposition of evolution of a dynamical reconfiguring network when foraging behaviour is used. Note that (a) and (c) show regions (for example in lower left corner, Cornwall) where part of the exploring growth front was trapped within the narrow confines of the map boundary.}
\label{model_superposition}
\end{figure}

By recording the paths of the particles during the evolution of the emergent transport networks is is possible to spatially approximate the variation seen in the \emph{Physarum}-graphs. Fig.~\ref{model_superposition} shows the effect of superimposing the trajectories of all particles during evolution as a single image. The most visited paths are shown as darker colours and the overall trajectory as a cloud-like mass around the nodes. Fig.~\ref{model_superposition}(a), formed in by the adaptation of a radially expanded population shows remnants of paths formed during the expansion process. Fig.~\ref{model_superposition}(b), formed by migratory extension shows a more space efficient search (corresponding somewhat to \emph{Physarum}-graph $\mathbf{P}_{10}$). The highly foraging superposition Fig.~\ref{model_superposition}(c) shows similarities to \emph{Physarum}-graph $\mathbf{P}_{5}$ and although there is significant rearrangement of the network (indicated by the fuzzy nature of the superposition), the central 'core' of the network (connecting the areas of West Midlands, Nottingham, Sheffield, West Yorkshire, Manchester and Liverpool) maintains its connectivity. This suggests that foraging is more significant in areas away from closely clustered sources of food, even when the cluster is far from the initial inoculation site.

\section{Imitating disasters}
\label{disasters}

When making experiments it is difficult to resist an impulse to imitate a large-scale disaster 
leading to contamination of  one of the urban areas spreading to the surrounding areas.
A disaster is imitated by placing a grain of salt in a substrate's loci corresponding 
to the urban area. Diffusion of sodium chloride in the substrate imitates progressive 
contamination of surrounding areas making them temporarily uninhabitable.  

\begin{figure}
\centering
\subfigure[$t=$32~h]{\includegraphics[width=0.32\textwidth]{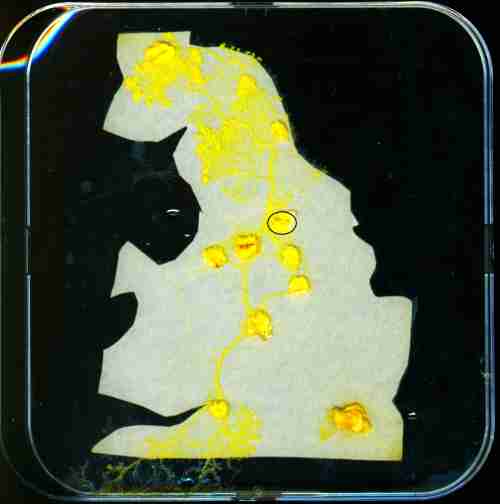}}
\subfigure[$t=$44~h]{\includegraphics[width=0.32\textwidth]{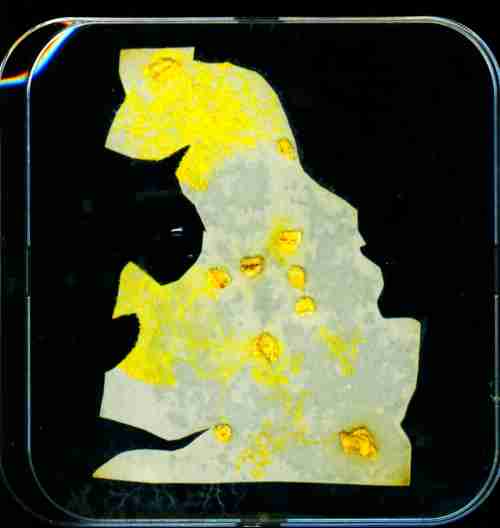}}
\subfigure[$t=$59~h]{\includegraphics[width=0.32\textwidth]{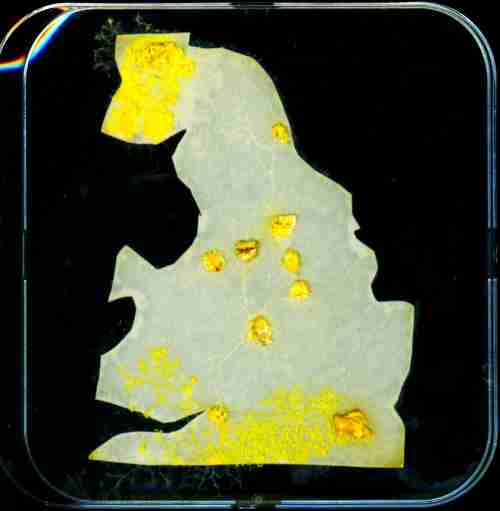}}
\subfigure[$t=$32~h]{\includegraphics[width=0.32\textwidth]{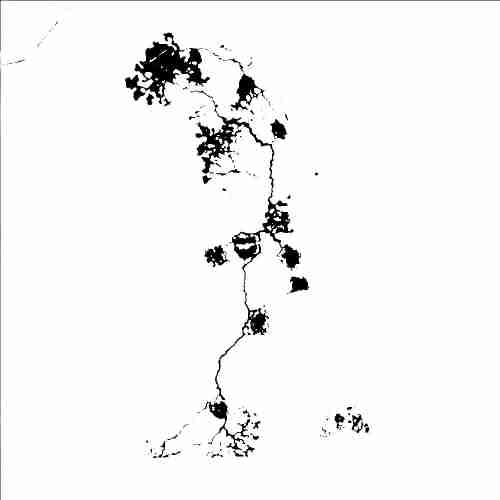}}
\subfigure[$t=$44~h]{\includegraphics[width=0.32\textwidth]{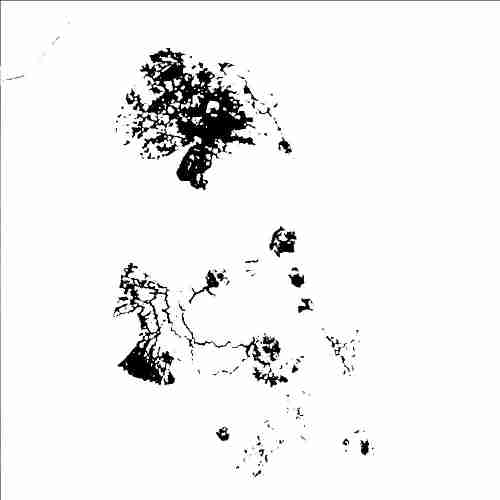}}
\subfigure[$t=$59~h]{\includegraphics[width=0.32\textwidth]{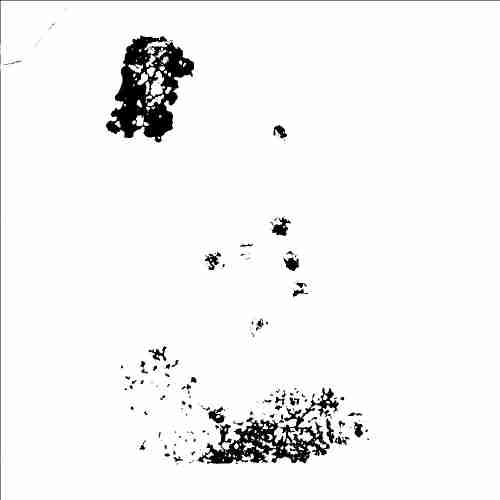}}
\caption{Plasmodium's response to disaster in, and subsequent contamination of 
West Yorkshire area, filter paper substrate: (a)--(c)~scanned images, (d)--(f)~binarized images, 
$\Theta=(200,200,20)$. Grain of salt placed (marked by circle in (a)) in West Yorkshire urban area at 
32~h of plasmodium development. Images (b)--(c) and (e)--(f) show 
the plasmodium's response to increased concentration of sodium chloride.}
\label{fr41a}
\end{figure}

\begin{figure}
\centering
\subfigure[$t=$37~h]{\includegraphics[width=0.32\textwidth]{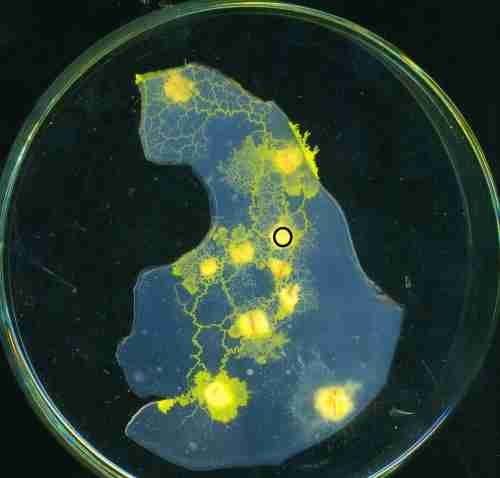}}
\subfigure[$t=$51~h]{\includegraphics[width=0.32\textwidth]{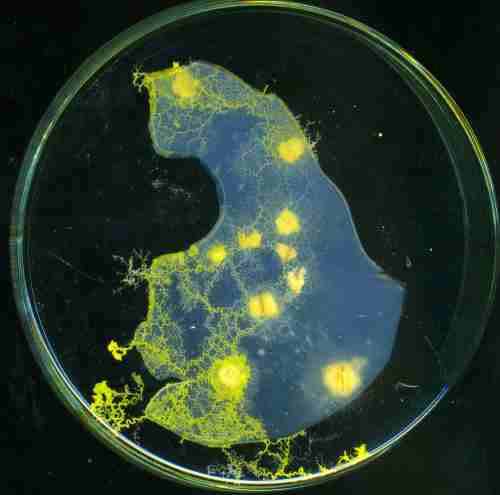}}
\subfigure[$t=$85~h]{\includegraphics[width=0.32\textwidth]{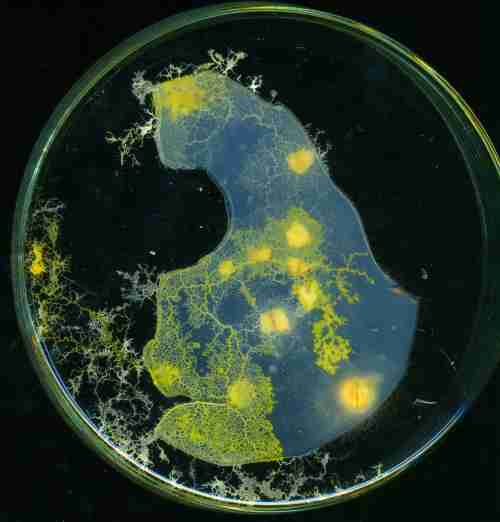}}
\subfigure[$t=$37~h]{\includegraphics[width=0.32\textwidth]{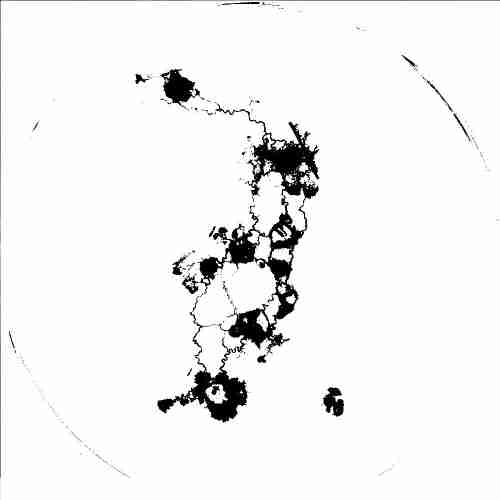}}
\subfigure[$t=$51~h]{\includegraphics[width=0.32\textwidth]{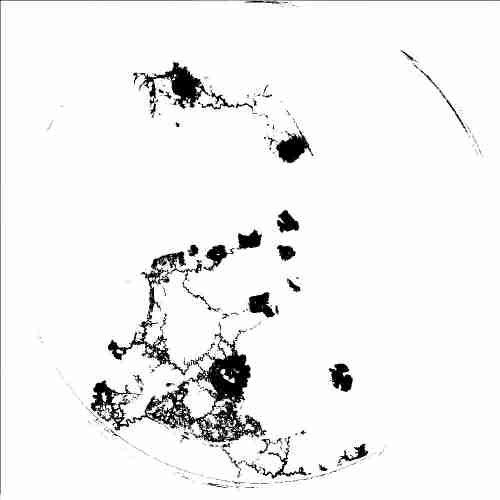}}
\subfigure[$t=$85~h]{\includegraphics[width=0.32\textwidth]{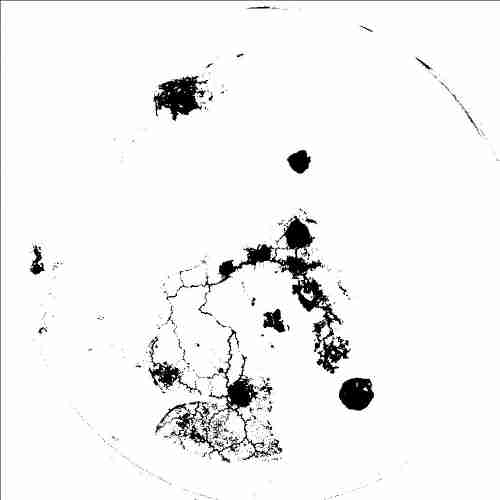}}
\caption{Plasmodium's response to disaster in, and subsequent contamination of 
West Yorkshire area, agar gel substrate: 
(a)--(c)~scanned images, (d)--(f)~binarized images, $\Theta=(200,200,20)$;
a grain of salt placed (marked by circle in (a)) in West Yorkshire urban area at 37~h of plasmodium development. Images (b)--(c) and (e)--(f) shows 
the plasmodium's reaction toward increased concentration of sodium chloride.}
\label{d5}
\end{figure}

\begin{figure}
\centering
\subfigure[$t=$32~h]{\includegraphics[width=0.32\textwidth]{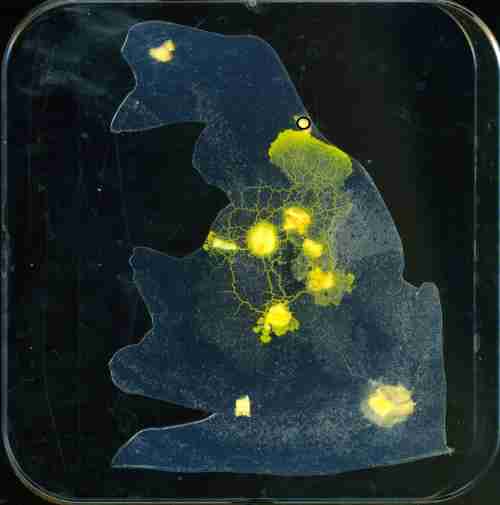}}
\subfigure[$t=$41~h]{\includegraphics[width=0.32\textwidth]{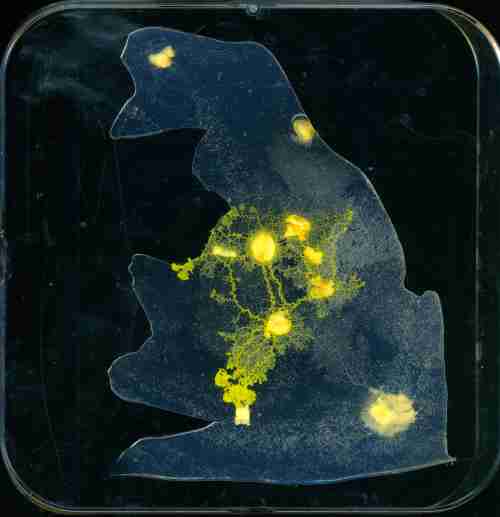}}
\subfigure[$t=$52~h]{\includegraphics[width=0.32\textwidth]{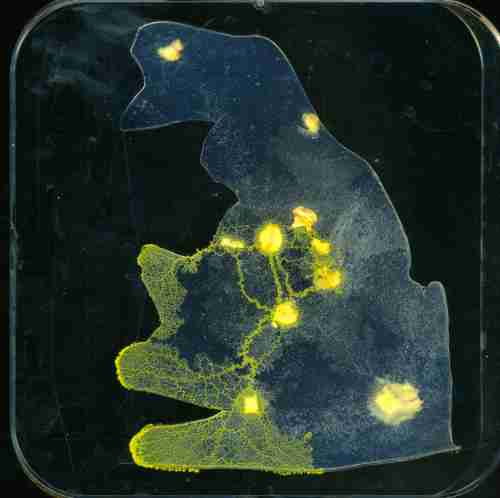}}
\subfigure[$t=$64~h]{\includegraphics[width=0.32\textwidth]{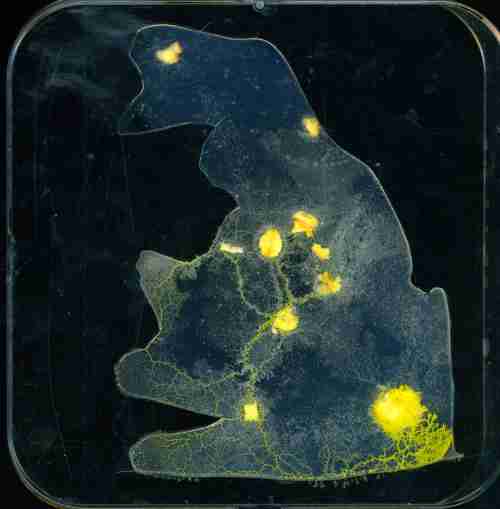}}
\subfigure[$t=$32~h]{\includegraphics[width=0.32\textwidth]{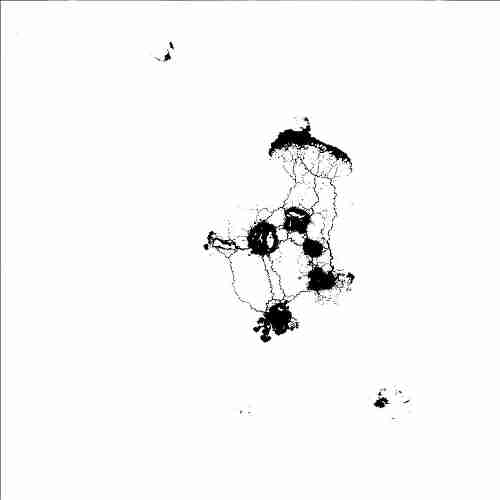}}
\subfigure[$t=$41~h]{\includegraphics[width=0.32\textwidth]{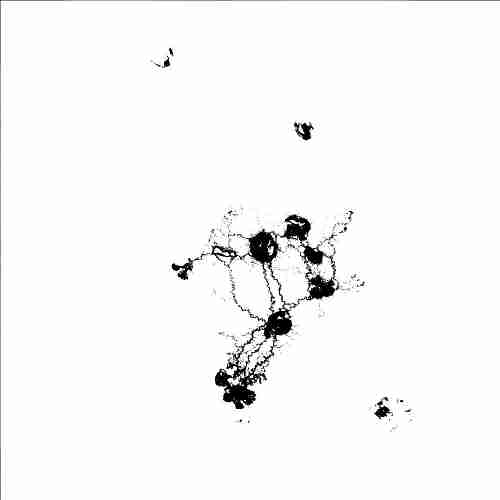}}
\subfigure[$t=$52~h]{\includegraphics[width=0.32\textwidth]{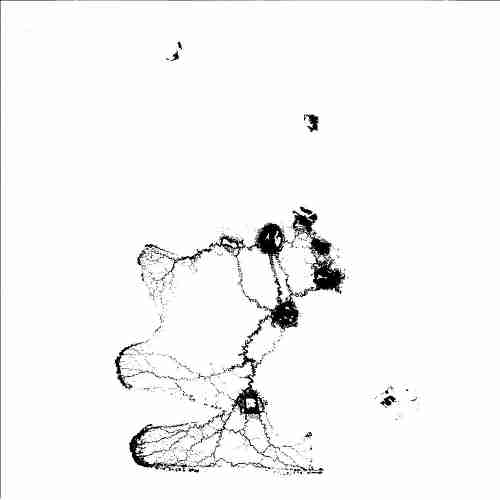}}
\subfigure[$t=$64~h]{\includegraphics[width=0.32\textwidth]{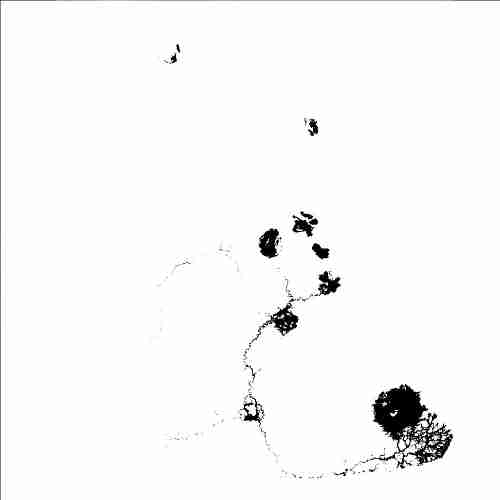}}
\caption{Plasmodium's response to disaster in, and subsequent contamination of 
Tyneside area, agar gel substrate, (a)--(d)~scanned images, (e)--(h)~binarized images, 
$\Theta=(200,200,20)$; 
a grain of salt placed (marked by circle in (a)) in Tyneside urban area at 32~h of 
plasmodium development. Images (b)--(c) and (e)--(f) shows the plasmodium's reaction 
toward increased concentration of sodium chloride.}
\label{b2}
\end{figure}

Here we consider three examples: disaster in and contamination of West Yorkshire urban 
area (Fig.~\ref{fr41a}, wet filter paper substrate, and Fig.~\ref{d5}, agar gel substrate) 
and Tyneside urban area (Fig.~\ref{b2}, agar gel substrate).

In experiment shown in Fig.~\ref{fr41a} a plasmodium network spanning most urban areas is formed 
32~h after inoculation of Greater London area with plasmodium. We place a salt crystal in Leeds (West Yorkshire 
urban area) (Fig.~\ref{fr41a}ad). In response to increased concentration of sodium chloride
the plasmodium abandons West Yorkshire area and disconnects the area from neighbouring Tyneside, Greater
Manchester and Sheffield urban areas (Fig.~\ref{fr41a}be). At the same time the plasmodium starts mass-exploration of 
Scotland and North Wales, and restores previously abandoned transport link with London (Fig.~\ref{fr41a}be). At around 27-30~h after
the `disaster' in West Yorkshire the plasmodium completes its evacuation from Midlands and North England and regroups itself
in North Scotland and South England (Fig.~\ref{fr41a}cf).   

Due to -- possibly -- a lower rate of sodium chloride diffusion in agar gel (comparing to wet filter paper), the plasmodium's 
response to imitated contamination of agar gel is less dramatic. As in the previous experiment, we wait until plasmodium forms 
a well-established protoplasmic network and then place a salt crystal in Leeds (Fig.~\ref{d5}ad). The plasmodium temporarily 
breaks all transport links leading to the contaminated zone (West Yorkshire), and increases exploratory activity in Wales (even developing pronounced protoplasmic routes in West Wales) and South-West England (Fig.~\ref{d5}be). One and a half day after contamination of West Yorkshire the plasmodium restores transport links with Leeds  (Fig.~\ref{d5}cf) and decrease its activity in Wales and South-West.

In the example shown in Fig.~\ref{b2} we strike urban area with contaminant before any transport link leading to the area is established. Plasmodium inoculated in Greater London area spans urban areas in Midlands with a protoplasmic network (Fig.~\ref{b2}ae). When the plasmodium's active zone approaches Tyneside area we place salt crystal in Newcastle (Fig.~\ref{b2}ae). In response to contamination the plasmodium abandons its attempt to colonize North England and Scotland
and guides its foraging activity towards the West and South, and 9~h after contamination of Tyneside the plasmodium reaches
Bristol (Fig.~\ref{b2}bf). The plasmodium increases its exploration of Wales and South-West England (Fig.~\ref{b2}cg) and, in 32~h after disaster strikes Tyneside, restores transport network between Midlands and London, 
this time via Bristol (Fig.~\ref{b2}cg).

\begin{finding}
Contamination of a single urban area stimulates exploration of uncolonized areas and leads to restoration of previously 
abandoned transport links. 
\end{finding}

\begin{figure}
\centering
\subfigure[$t=$ 2449, 2884, 3189, 3655, 3855 and 4372 scheduler steps]{\includegraphics[width=0.8\textwidth]{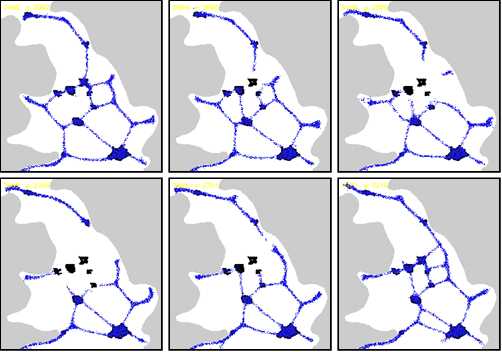}}
\subfigure[$t=$ 117, 199, 321, 443, 1190, 1539, 2812 and 15000 scheduler steps]{\includegraphics[width=0.95\textwidth]{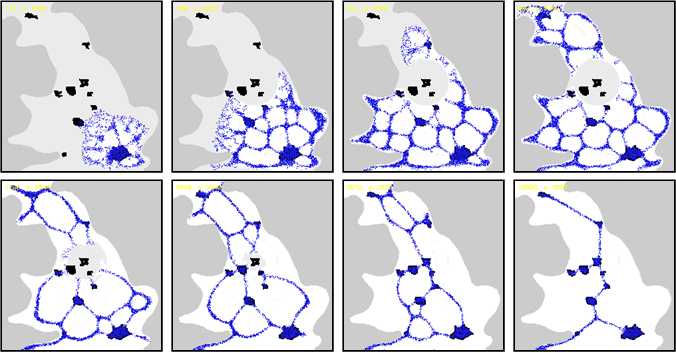}}
\caption{Effect of disastrous contamination in the West Yorkshire urban area after and prior to network formation. (a) Foraging behaviour resulted in connection of nodes. Simulation of contamination by Salt with increasing radius. At $t=$ 4228 contamination is removed and network reconnects with previously contaminated nodes. (b) Growth of simulated plasmodium in a nutrient-rich substrate where West Yorkshire urban area is already contaminated by salt (radius 37 pixels). At $t=$379 the growth front by-passes contaminated area. At $t=$1289 the contaminated area shrinks and the previously contaminated nodes are added to the network.}
\label{model_contaminated_leeds}
\end{figure}

To approximate contamination using the particle model we introduced simulated salt to the West Yorkshire urban area into a pre-existing foraging transport network (Fig.~\ref{model_contaminated_leeds} (a)). The effect of particle exposure to contaminated areas resulted in the removal of those particles from the environment. The contaminated region increased in radius and resulted in disconnection of contaminated nodes from the network. Although this resulted in increased foraging in nearby areas, the results were not appreciably different from the general foraging behaviour of the particle approximation and certainly not as dramatic as observed in the organism itself. When the contamination was removed the foraging behaviour resulted in reconnection of the transport network. In Fig.~\ref{model_contaminated_leeds} (b) we simulated the growth of a plasmodium from the London urban area in a nutrient-rich substrate where the West Yorkshire urban area was already contaminated by salt. The contaminated area 'killed' any particles entering the affected region and the growth wave by-passed the region. At $t=$ 1289 the contamination was removed and the growth front began to extend into the West Yorkshire area, reconnecting the network. After the network was fully connected, minimisation of the network occurred in the usual manner.

\section{Discussion}
\label{discussion}

We undertook scoping laboratory and simulation experiments on bio-inspired road planning. We represented the ten most populated UK urban areas by source of nutrients, inoculated plasmodium of \emph{Physarum polycephalum} in one of the areas and analysed dynamics of colonizing areas by the plasmodium. We performed the same spatial representation of urban areas in a spatially represented particle approximation of \emph{Physarum} under three different environmental conditions: nutrient-rich background substrate, high quality nutrients in a strongly diffusive environment, and a poor quality environment with foraging behaviour. We studied space-time dynamics of spanning the urban areas by the plasmodium's network of protoplasmic tubes and demonstrated that the plasmodium transport network sufficiently well matches the topology of the existing man-made motorway networks with two exclusions: Motorway M4 (Bristol--London) rarely occurs in plasmodium networks, and the route M6/M74 is entirely absent. Very similar network patterns were observed in the particle approximation which also succeeded in minimising the transport network configuration in nutrient-rich environments. As a `by-product' of the experiments we provided an insight into bio-inspired response to disastrous contamination of urban areas. Two main components of the response are exploring non-colonised territories and restoration of abandoned transport links. The approximation of disastrous contamination in the particle model reflected the temporary destruction of the transport networks, their reconfiguration after the contamination had passed, but did not show evidence of dramatic shifts in foraging behaviour when exposed to the contaminant. We believe that the lack of dramatic response of the particle collective is because the oscillatory behaviour seen in \emph{Physarum} was not included in this instance of the particle collective. We aim to integrate growth behaviours with oscillatory behaviours in future works. 

Our experiments did not take terrain into account. This may explain the `anomalous' situation with plasmodium not imitating route M6/M74 but developing a transport link directly from Newcastle to Glasgow. Terrain based experiments on \emph{Physarum} road planning may form a subject for further studies.

\section{Acknowledgement}

The work was partially supported by the Leverhulme Trust research grant F/00577/1 ``Mould intelligence: biological amorphous robots''.

\end{document}